\documentclass[prb,english,aps,prb,preprint,superscriptaddress,showpacs,showkeys]{revtex4-1}
\usepackage{bm}
\usepackage{amsmath}
\usepackage{amssymb}
\usepackage{graphicx}

\usepackage{babel}
\begin{document}

\title{Electron-electron interaction in graphene at finite Fermi energy.}

\author{A.I. Milstein}

\email{A.I.Milstein@inp.nsk.su}

\author{I.S. Terekhov}

\email{I.S.Terekhov@inp.nsk.su}

\affiliation{Budker Institute of Nuclear Physics of SB RAS, Novosibirsk, 630090
Russia}

\begin{abstract}
The wave equation  describing the interaction of two electrons in graphene at arbitrary value of the Fermi energy $E_F$ is derived. For the solutions of this equation, we have found the explicit forms of the density and the current  which obey the continuity equation.
We have traced  the evolution of the wave packet during a scattering process. It is shown that the long-leaving localized quasi-stationary peak may appear at  $E_F<0$ . Then this peak  decays   into a set of wave packets following each other. At $t\rightarrow\infty$  a total norm of all outgoing wave packets equals to that of incoming wave packet.  At $E_F=0$ the localized state does not appear.
For $E_F<0$ there is an infinite set of the localized solutions with the  finite  norms.

\end{abstract}
\maketitle

\section{Introduction}
Investigation of electron-electron scattering in graphene is very important task because the results of this investigation may explain a high mobility of the charge carriers \cite{Novoselov2004}. 
Nowadays there is a set of  theoretical publications  devoted to this subject \cite{Sabio2010,LMT2012},  see also reviews \cite{Kotov2011,DasSarma2011}.  
It is well established now that the low-energy single electron dynamics
in graphene is described by the massless two-component Dirac equation
\cite{Wallace1947,McClure1956,Gonzalez1993,Gonzalez1994,Novoselov2004}
\begin{gather*}
i\hbar\partial_{t}\psi\left(t,\bm{r}\right)=v_{F}\bm{\sigma}\cdot\hat{\bm{p}}\,\psi\left(t,\bm{r}\right),
\end{gather*}
where $v_{F}$ is the Fermi velocity, $\hat{\bm{p}}=-i\hbar\bm{\nabla}$,
and $\bm{\sigma}=\left(\sigma_{x},\sigma_{y}\right)$ are the Pauli
matrices acting on the pseudospin variables. Below we set $\hbar=v_{F}=1$.
A pair of non-interacting electrons can be described by the equation
\begin{eqnarray}\label{free}
&&i\partial_{t}\psi\left(\mathbf{r}_{1},\mathbf{r}_{2},t\right)=  \hat{H}_{0}\psi\left(\bm{r}_{1},\bm{r}_{2},t\right)\,,\nonumber\\
&&\hat{H}_{0} =\bm{\sigma}_{1}\cdot\hat{\bm{p}}_{1}+\bm{\sigma}_{2}\cdot\hat{\bm{p}}_{2}\,,
\end{eqnarray}
where $\psi\left(\mathbf{r}_{1},\mathbf{r}_{2},t\right)$ is the wave
function depending on the coordinates and pseudospin
variables of both electrons. There is a nontrivial problem of  generalization of  Eq. \eqref{free}
to the case of interacting electrons. This problem arises because of  necessity to take into account
the electron-hole interaction, i.e., the interaction of electrons above Fermi surface with electrons below Fermi surface. In quantum electrodynamics this interaction  is  account for by means of the Dyson-Schwinger equation, see, e.g., Ref.\cite{Berestetski1982,IZ}.
For massless electrons in graphene, an account for the electron-hole interaction may qualitatively change the properties of the electron-electron interaction. The approach
based on the Bethe-Salpeter equation (reduction of  the Dyson-Schwinger equation) was used in Ref. \cite{Gamayun2009,Gamayun2010} at the investigation of the electron-hole interaction in graphene.

In Refs. \cite{Sabio2010, LMT2012} the electron-electron interaction has been studied using the equation \eqref{free} with the  replacement $\hat{H}_{0}\rightarrow\hat{H}_{0}+V\left(r\right)$,
where $V\left(r\right)=V(\left|\bm{r}_{1}-\bm{r}_{2}\right|)$ is the
electron-electron interaction potential. This means that the electron-hole interaction has been neglected. In the frame of zero total momentum the corresponding wave equation reads
\begin{align}\label{za}
& i\partial_{t}\psi\left(t,\mathbf{r}\right)=\hat{H}\psi\left(t,\mathbf{r}\right),\nonumber\\
& \hat{H}=\left(\boldsymbol{\sigma}_{1}-\boldsymbol{\sigma}_{2}\right)\cdot\hat{\mathbf{p}}+
V\left(r\right)\,.
\end{align}
In Ref.\cite{port2017} the stationary normalized solutions of Eq.~\eqref{za} with zero energy are found. In Refs.\cite{Sabio2010, LMT2012} it is shown that the solutions of the wave equation \eqref{za} have unusual properties. In the present paper we investigate the impact of the electron-hole interaction on these properties. We show that the solutions found in Refs.\cite{Sabio2010, LMT2012} correspond to the Fermi energy $E_f\rightarrow -\infty$. We also show that  the electron-electron interaction at $E_f=0$ has completely different properties  compared to that found in  Refs.\cite{Sabio2010, LMT2012}. 
A transformation of the solutions with increasing $E_f$  from $-\infty$ to zero is  traced.

\section{General properties of the model}
We start our consideration with the derivation of the wave equation for two interacting electrons with an  account for  the electron-hole interaction. 

The Bethe-Salpeter equation for the two-body  function $\Phi(\varepsilon_1,\bm p_1|\varepsilon_2,\bm p_2) $, see, e.g., Ref.\cite{IZ}, has the form 
\begin{align}\label{DS1}
&\Phi(\varepsilon_1,\bm p_1|\varepsilon_2,\bm p_2)=iG(\varepsilon_1,\bm p_1)G(\varepsilon_2,\bm p_2)
 \int \dfrac{d\bm q\, d\omega}{(2\pi)^3}{\widetilde V(q)}\Phi(\varepsilon_1+\omega,\bm p_1+\bm q|\varepsilon_2-\omega,\bm p_2-\bm q)\,,\nonumber\\
&G(\varepsilon_i,\bm p_i)=\frac{1}{\varepsilon_i-\bm{\sigma}_{i}\cdot\bm{p}_{i}+i0\,{\rm sgn}(\varepsilon_i-E_F)}\,,
\end{align}
where $\widetilde V(q)$ is the Fourier transform of the potential $V(r)$, $G(\varepsilon, \bm p)$ is the one-particle  Green's function, and $E_f$ is the Fermi energy.
We make the substitution $\varepsilon_1= E/2+\Omega$, $\varepsilon_2= E/2-\Omega$ and take the integral over $\Omega$ in the both sides of Eq.~\eqref{DS1}. We have
\begin{align}\label{DS2}
&\Xi(E,\bm p_1,\bm p_2)=i\int \dfrac{d\Omega}{2\pi} G\left(\frac{1}{2}E+\Omega,\bm p_1\right)G\left(\frac{1}{2}E-\Omega,\bm p_2\right)\nonumber\\
&\times \int \dfrac{d\bm q}{(2\pi)^2}{\widetilde V(q)}\,\Xi(E,\bm p_1+\bm q,\bm p_2-\bm q)\,,
\end{align}
where
$$\Xi(E,\bm p_1,\bm p_2)=\int \frac{d\Omega}{2\pi}\Phi\left(\dfrac{1}{2}E+\Omega,\bm p_1|\dfrac{1}{2}E-\Omega,\bm p_2\right).$$
Performing  in Eq.~\eqref{DS2} the integration  over $\Omega$ we finally  obtain the following equation for the wave function $\Xi(E,\bm p_1,\bm p_2)$:
\begin{align}\label{DS3}
&(E-\bm{\sigma_1}\cdot\bm p_1-\bm{\sigma_2}\cdot\bm p_2)\Xi(E,\bm p_1,\bm p_2)=\frac{1}{2}\Big \{\bm{\sigma_1}\cdot\bm n_1\vartheta(p_1-|E_F|)+\bm{\sigma_2}\cdot\bm n_2\vartheta(p_2-|E_F|)\nonumber\\
&-\mbox{sgn}(E_F)[\vartheta(|E_F|-p_1)+\vartheta(|E_F|-p_2)]\Big\}
\int \dfrac{d\bm q}{(2\pi)^2}{\widetilde V(q)}\Xi(E,\bm p_1+\bm q,\bm p_2-\bm q)\,,
\end{align}
where $\bm n_1=\bm p_1/p_1$ and $\bm n_2=\bm p_2/p_2$. This equation is valid for any $\bm p_1$, $\bm p_2$, and $E_f$. Below we consider the most interesting case $\bm p_1=-\bm p_2=\bm p$ and $\varepsilon_F\equiv-E_F\geq 0$ (Fermi surface is below or coincides with the Dirac point). Then  we have
\begin{align}\label{DS4}
&(E-\bm{\Sigma}\cdot\bm p)\chi(E,\bm p)
=R\, \int \dfrac{d\bm q}{(2\pi)^2}{\widetilde V(q)}\chi(E,\bm p+\bm q)\,,\nonumber\\
&R=\frac{1}{2}\bm{\Sigma}\cdot\bm n\,\vartheta(p-\varepsilon_F)
+\vartheta(\varepsilon_F-p)\,,
\end{align}
where $\bm n=\bm p/p$, $\bm\Sigma=\bm\sigma_1-\bm\sigma_2$, and $\chi(E,\bm p)=\Xi(E,\bm p,-\bm p)$.
As should be, in the limit  $\varepsilon_F\rightarrow\infty$ ($E_F\rightarrow -\infty$) the equation \eqref{DS4} transfers in Eq.~\eqref{za} written in the momentum representation.

To have  the physical interpretation of the wave equation \eqref{DS4}, it is necessary to derive the continuity equation. For this purpose, let us introduce the projector operators 
\begin{equation}
\Lambda^{\pm\pm}(\bm p)=\frac{1}{4}(1\pm \bm\sigma_1\cdot\bm n)(1\mp \bm\sigma_2\cdot\bm n)\,,
\end{equation}
and  the operators
\begin{align}
&L_{1}(\bm p)=\vartheta(p-\varepsilon_F)\Lambda^{++}(\bm p)+\vartheta(\varepsilon_F-p)\,,\quad L_{2}(\bm p)=\vartheta(p-\varepsilon_F)\Lambda^{--}(\bm p)\,,\nonumber\\
&L_{3}(\bm p)=\vartheta(p-\varepsilon_F)\Lambda^{+-}(\bm p)\,,\quad L_{4}(\bm p)=\vartheta(p-\varepsilon_F)\Lambda^{-+}(\bm p)\,.
\end{align}
The  latter operators are also projectors, i.e., $L_iL_j=\delta_{ij}L_i$ and $\sum_{i=1}^4L_i=1$. 
We also introduce the functions
\begin{align}
&\chi_+(E,\bm p)=L_{1}(\bm p)\chi(E,\bm p)\,,\quad\chi_-(E,\bm p)=L_{2}(\bm p)\chi(E,\bm p)\,,\nonumber\\
&\chi_{+-}(E,\bm p)=L_{3}(\bm p)\chi(E,\bm p)\,,\quad\chi_{-+}(E,\bm p)=L_{4}(\bm p)\chi(E,\bm p)\,.
\end{align}
Note that $\chi=\chi_++\chi_-+\chi_{+-}+\chi_{-+}$.
It follows from Eq.~\eqref{DS4} that $\chi_{+-}(E,\bm p)=\chi_{-+}(E,\bm p)=0$, so that 
$\chi=\chi_++\chi_-$. The functions  
$\chi_{+}(E,\bm p)$ and $\chi_{-}(E,\bm p)$ obey the equations
\begin{align}\label{DS5}
&(E-\bm{\Sigma}\cdot\bm p)\chi_+(E,\bm p)=L_1(\bm p)\, \int \dfrac{d\bm q}{(2\pi)^2}{\widetilde V(q)}\chi(E,\bm p+\bm q)\,,\nonumber\\
&(E-\bm{\Sigma}\cdot\bm p)\chi_-(E,\bm p)=-L_2(\bm p)\, \int \dfrac{d\bm q}{(2\pi)^2}{\widetilde V(q)}\chi(E,\bm p+\bm q)\,.
\end{align}
The time dependent wave  functions written in the coordinate space
($\chi_{\pm}(E,\bm p)\rightarrow \psi_{\pm}(t,\bm{r})$) obey the equation: 
\begin{align}\label{eqpsi}
&(i\partial_{t}-\bm{\Sigma}\cdot\hat{\bm{p}})\,\psi_{\pm}(t,\bm{r})=\pm\int d\bm r' Q_{\pm}(\bm r-\bm r') V(r')\psi(t,\bm{r}')\,,\nonumber\\
&Q_+(\bm r)=f_0+f_1+f_2\,,\quad Q_-(\bm r)=-f_1+f_2\,,\nonumber\\
&f_0=\frac{yJ_1(y)}{2\pi r^2}\,,\quad f_1=i\frac{\bm\Sigma\cdot\bm y}{8\pi r^2y}\left[yJ_0(y)+1-\int_0^yJ_0(x)\,dx\right]\nonumber\\
&f_2=\frac{1}{4}\left[\delta(\bm{r})-\dfrac{yJ_1(y)}{2\pi r^2}\right]\left(1-\frac{\bm{\sigma}_1\cdot\bm{\sigma}_2}{2}\right)\nonumber\\
&+\dfrac{2(\bm{\sigma}_1\cdot\bm{y}) (\bm{\sigma}_2\cdot\bm{y})-y^2\bm{\sigma}_1\cdot\bm{\sigma}_2}{16\pi r^2y^2}[2J_0(y)+yJ_1(y)]\,,
\end{align}
where $\bm y=\varepsilon_F\bm r$, $J_n(x)$ are the Bessel functions of the first kind, and $\psi(t,\bm r)=\psi_+(t,\bm r)+\psi_-(t,\bm r)$. The operators $Q_+(\bm r)$ and $Q_-(\bm r)$ are the projector operators in the coordinate space corresponding to the projector operators $L_1(\bm p)$ and $L_2(\bm p)$ in the momentum space.
It follows from Eq.~\eqref{eqpsi} that the wave function $\psi(t,\bm r)$ obeys the equation
\begin{align}\label{eqpsi1}
&(i\partial_{t}-\bm{\Sigma}\cdot\hat{\bm{p}})\,\psi(t,\bm{r})=\int d\bm r'[ Q_{+}(\bm r-\bm r')-Q_{-}(\bm r-\bm r')] V(r')\psi(t,\bm{r}')\,.
\end{align}

Using \eqref{eqpsi} we find 
\begin{align}\label{eqcon}
&\partial_{t}\rho(t,\bm r)+\mbox{div} \bm J(t,\bm r)+F(t,\bm r)=0\,,\nonumber\\
&\rho=\psi^+_+\psi_+-\psi^+_-\psi_-\,,\quad \bm J=\psi^+_+\bm{\Sigma}\psi_+-\psi^+_-\bm{\Sigma}\psi_-\,,\nonumber\\
&F=2\,\mbox{Im}\int d\bm r'\,V(r')\left[\psi_+^+(t,\bm r)Q_+(\bm r-\bm r')+
\psi_-^+(t,\bm r)Q_-(\bm r-\bm r')\right]\psi(t,\bm r')\,.
\end{align}
Since $Q_+$ and $Q_-$ are the hermitian projector operators, then
\begin{equation}
\int d\bm r\, F(t,\bm r)= 2\,\mbox{Im}\int d\bm r'\,V(r')\psi^+(t,\bm r')\psi(t,\bm r')=0\,.
\end{equation}
Therefore $\int d\bm r\,\rho(t,\bm r)$ is time-independent. The equation \eqref{eqcon} may be written in the conventional form
\begin{align}\label{eqconu}
&\partial_{t}\rho(t,\bm r)+\mbox{div} \bm J_{tot}(t,\bm r)=0\,,\nonumber\\
&\bm J_{tot}=\bm J+\dfrac{1}{2\pi}\int d\bm r'\,\dfrac{\bm r-\bm r'}{|\bm r-\bm r'|^2}\,F(\bm r') \,.
\end{align}
The validity of the continuity equation allows us to treat the function $\psi(t,\bm r)$ as a wave function of two electrons and the quantity $e\rho(t,\bm r)$ as  a local charge density ($e$ is the electron charge). Note that, generally speaking,  $\rho(t,\bm r)$ is not positive, but the charge density should not be  positive. 

\section{Time evolution of the wave packets.}
To investigate a time evolution of the wave packets, we write the equation for the wave function $\chi(t,\bm p)$ in the form
\begin{align}\label{DS4t}
&(i\partial_t-\bm{\Sigma}\cdot\bm p)\chi(t,\bm p)
=R\, \int \dfrac{d\bm q}{(2\pi)^2}{\widetilde V(q)}\chi(t,\bm p+\bm q)\,,
\end{align}
where $R$ is given in Eq.~\eqref{DS4}.  Then we represent the function $\chi(t,\bm p)$ as
\begin{align}\label{ex}
&\chi(t,\bm p)= \sum_m c_m\chi_m(t,p,\varphi)\,,\nonumber\\
& \chi_m(t,p,\varphi)=
e^{im\varphi}\Bigl[a_{00}\left(t,p\right)\left|0,0\right\rangle +e^{-i\varphi}a_{11}\left(t,p\right)\left|1,1\right\rangle \nonumber\\
&  +e^{i\varphi}a_{1-1}\left(t,p\right)\left|1,-1\right\rangle+
  g\left(t,p\right)\left|1,0\right\rangle \Bigr]\,,
\end{align}
where $c_m$ are some constants, $ \chi_m(t,p,\varphi)$ are the eigenfunctions of the operator $$J^z=T^z-i\partial_{\varphi}$$
with the eigenvalue $m$, where  $\bm T=(\bm\sigma_{1}+\bm\sigma_{2})/2$; $|1,k\rangle$ and $|0,0\rangle$ are the eigenfunctions of the operator 
$\bm T^2$ and $T^z$. Let us consider the evolution of the function $ \chi_m(t,p,\varphi)$. It is convenient to pass from the functions $a_{ij}$ to the functions $f,\,h,$ and $d$:
\begin{equation}
f=\frac{a_{11}+a_{1-1}}{\sqrt{2}}\,,\quad h=\frac{a_{11}-a_{1-1}}{\sqrt{2}}\,,\quad d=a_{00}.\label{eq:fhdg definition}
\end{equation}
These functions  obey the system of integro-differential equations
\begin{align}\label{geq}
&i\partial_t f=\vartheta(\varepsilon_F-p)\left[{\hat U}_+f+{\hat U}_-h\right]\,,\nonumber\\
&i\partial_t h=-2pd+\vartheta(\varepsilon_F-p)\left[{\hat U}_-f+{\hat U}_+h\right]-\vartheta(p-\varepsilon_F){\hat U}_0d\,,\nonumber\\
&i\partial_t d=-2ph-\vartheta(p-\varepsilon_F)\left[{\hat U}_-f+{\hat U}_+h\right]+\vartheta(\varepsilon_F-p){\hat U}_0d\,,\nonumber\\
&i\partial_t g=\vartheta(\varepsilon_F-p){\hat U}_0g\,.
\end{align} 
Here the following notations are used
\begin{align}
&{\hat U}_\pm H=\frac{1}{2}\int_0^\infty\,dk\, k[V_{m-1}(p,k)\pm V_{m+1}(p,k)]H(k)\,,\quad
{\hat U}_0 H=\int_0^\infty\,dk\, kV_{m}(p,k)H(k)\,,\nonumber\\
&V_{m}(p,k)=\int_0^\infty\,dr\, rV(r)\,J_m(pr)J_m(kr)\,,
\end{align}  
where $H(k)$ is an arbitrary function. 
In the limiting case $\varepsilon_F\rightarrow\infty$, we have
\begin{align}\label{geqold}
&i\partial_t f={\hat U}_+f+{\hat U}_-h\,,\quad
i\partial_t h=-2pd+{\hat U}_-f+{\hat U}_+h\,,\nonumber\\
&i\partial_t d=-2ph+{\hat U}_0d\,,\quad
i\partial_t g={\hat U}_0g\,.
\end{align} 
This system of equations have been investigated in detail in Ref.~\cite{LMT2012} in coordinate space.
It turns out that the solutions at $m\neq 0$ have very unusual properties. 
Namely, the time evolution of the wave packet, corresponding to the scattering problem setup,
leads to the appearance of the localized state at large time. This is because the first equation in \eqref{geqold}, written in a coordinate space, does not contain  derivatives over a  spatial variable $r$ and reduces  to the equation of constraint.  Below we trace  whether the localized states  survive at a finite value of $\varepsilon_F$.

Note that the system \eqref{geq} at $m=0$ does not reveal the unusual properties at any $\varepsilon_F$. This statement can be explained as follows. For $m=0$ we have $U_-=0$ and the system \eqref{geq} reduces to  
\begin{align}\label{geqm0}
&i\partial_t f=\vartheta(\varepsilon_F-p){\hat U}_+f\,,\nonumber\\
&i\partial_t h=-2pd+\vartheta(\varepsilon_F-p){\hat U}_+h-\vartheta(p-\varepsilon_F){\hat U}_0d\,,\nonumber\\
&i\partial_t d=-2ph-\vartheta(p-\varepsilon_F){\hat U}_+h+\vartheta(\varepsilon_F-p){\hat U}_0d\,,\nonumber\\
&i\partial_t g=\vartheta(\varepsilon_F-p){\hat U}_0g\,.
\end{align} 
It is possible to show that the functions $f$ and $g$ at $m=0$ tend to zero at $r\rightarrow\infty$ faster than $1/\sqrt{r}$, so that  they are irrelevant to the scattering problem. Then we have a coupled  system of equation for the functions $h$ and $d$ without any constraints and without any unusual properties. This is why for the scattering problem only the cases $m\neq 0$  is investigated below.

Let us consider  the  convergent at $t\rightarrow -\infty$ wave packet, corresponding to $m\neq 0$,  with the average energy $E$  and some width $\delta l$ (the energy spread is $\delta E\sim 1/\delta l\ll E$) scattered  on the potential 
\begin{equation}\label{pot}
V(r)=u_0\exp[-r^2/a^2] 
\end{equation}
with $u_0>0$. We also assume that $\delta E\ll u_0$. A goal of this section is to trace the evolution of the density  $\rho(t,r)$. To calculate this function, we firstly find  the solutions  of Eq.~\eqref{geq} and then pass to the coordinate space making the Fourier transform. As a result we come to the expression for the density  $\rho(t,r)=\rho_+(t,r)-\rho_-(t,r)$ corresponding to the function $\chi_m(t, p, \phi)$: 
\begin{align}\label{rho}
&\rho_{\pm}(t,r)=|a_\pm|^2 + |b_\pm|^2 + 2|c_\pm|^2\,,\nonumber\\
&\left(
\begin{array}{c}
a_+\\
b_+\\
c_+
\end{array}\right)=\int_{\varepsilon_F}^{\infty}dp\,p(h-d)
\left(
\begin{array}{c}
J_{m-1}\\
J_{m+1}\\
J_{m}
\end{array}\right)+2\int_0^{\varepsilon_F}dp\,p
\left(
\begin{array}{c}
(h+f)J_{m-1}\\
(h-f)J_{m+1}\\
-dJ_{m}
\end{array}\right)
\,, \nonumber\\
&\left(
\begin{array}{c}
a_-\\
b_-\\
c_-
\end{array}\right)=\int_{\varepsilon_F}^{\infty}dp\,p(h+d)
\left(
\begin{array}{c}
J_{m-1}\\
J_{m+1}\\
J_{m}
\end{array}\right)
\,,
\end{align}
where the arguments of all Bessel functions are $pr$ and the functions $f$, $h$, and $d$ are the solutions of Eq.~\eqref{geq}. Note that the function $g(t,r)$ is irrelevant to the scattering problem at any $m$, and we have omitted it in Eq.~\eqref{rho}.  

\begin{figure}[h]
\includegraphics[width=0.45\linewidth,clip]{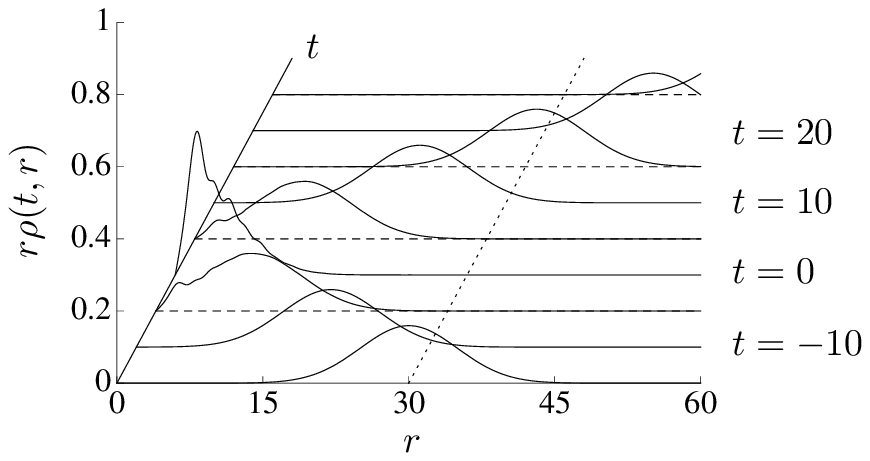}
\includegraphics[width=0.45\linewidth,clip]{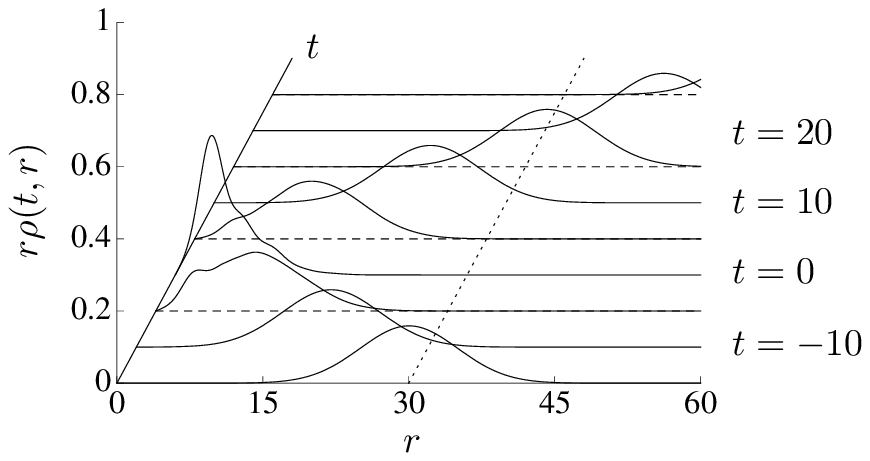}
\caption{Time evolution of the function $r\rho(t,r)$ for the wave packet with the parameters  $m=1$, $\delta E=0.1$, $E=4$ (left) and  $E=2$ (right) scattered on the potential \eqref{pot}  with the parameters $u_0=3$, $a=2$; $\varepsilon_F=0$.\label{ef0}}
\end{figure}

Below we analyze the  process at a few values of $\varepsilon_F$. 
We start with  the case  $\varepsilon_F=0$ when the system \eqref{geq} reduces to 
\begin{align}\label{geq0}
&i\partial_t f=0\,,\quad i\partial_t g=0\,,\nonumber\\
&i\partial_t h=-2pd-{\hat U}_0d\,,\quad 
i\partial_t d=-2ph-{\hat U}_-f-{\hat U}_+h\,.
\end{align} 
It is seen that the functions $f$ and $g$ are time-independent, so that they are irrelevant to the scattering process, and we can set them to be equal to zero. Thus, we have a coupled system of equations
\begin{align}\label{geq01}
i\partial_t h=-2pd-{\hat U}_0d\,,\quad 
i\partial_t d=-2ph-{\hat U}_+h\,.
\end{align} 
In contrast to Eqs.~\eqref{geqold}, the system \eqref{geq01} does not contain any constraints, so that the localized states do not appear during a scattering process. To illustrate this statement,
we show in Fig.~\ref{ef0} the time evolution of the function $r\rho(t,r)$, see Eq.~\eqref{eqcon}, for  scattering of the wave packet  with the parameters $m=1$,  $\delta E=0.1$, $ E=4$ (left picture) and  $E=2$ (right picture) on the potential with  the parameters  $u_0=3$ and $a=2$. 
It is seen from Fig.~\ref{ef0}
that localized states have not appeared, the shape  of the outgoing wave is  the same as that of the incoming one, and $\int dr\,r\,\rho(-\infty,r)=\int dr\,r\,\rho(\infty,r)$. This statement is valid for both cases
$E>u_0$ and $E<u_0$.

\begin{figure}
\includegraphics[width=0.45\linewidth,clip]{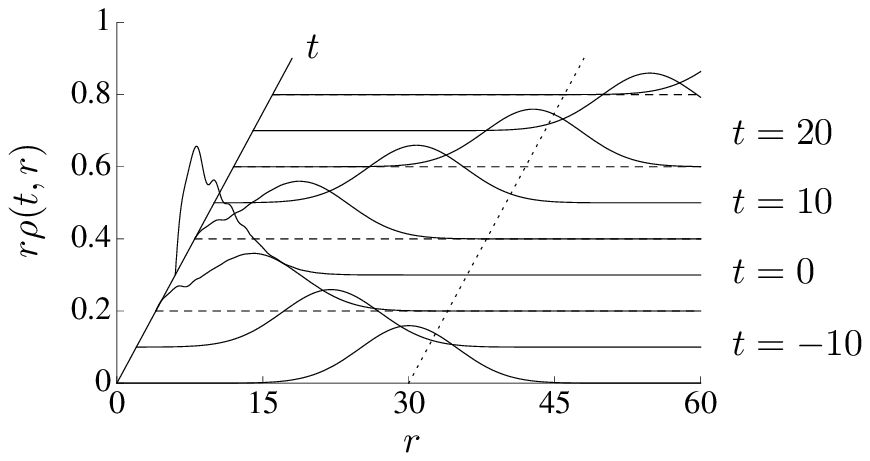}
\includegraphics[width=0.45\linewidth,clip]{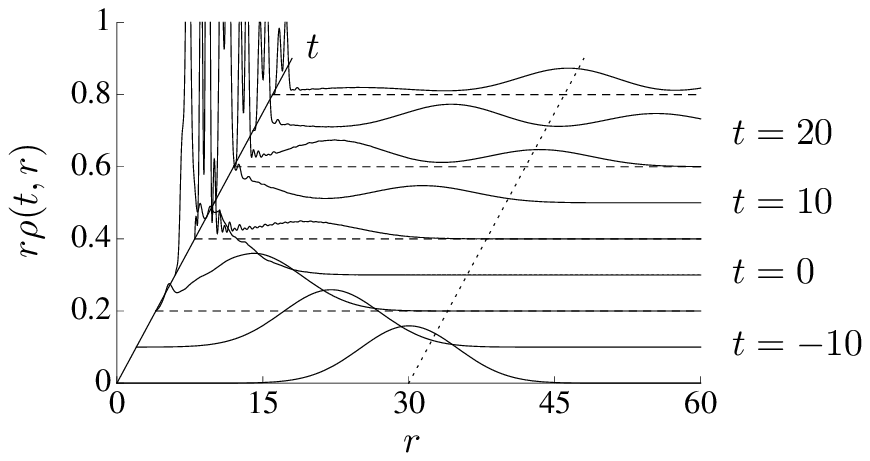}
\caption{Same as in Fig.\ref{ef0} but for $\varepsilon_F=6$.\label{ef6}}
\end{figure}

Then we pass to the case $\varepsilon_F=6$ at the same parameters of the incoming wave packet and the  potential. The evolution of the function $r\rho(t,r)$ is shown in Fig.~\ref{ef6}. This picture is typical for the evolution of the wave packet at non-zero but finite $\varepsilon_F$.
For $E>u_0$ (left picture) we have only one  outgoing wave with the same shape as the incoming wave  and with the same norm. For $E<u_0$ (right picture) the situation is completely different.
When the initial wave came to the nonzero potential region, the long-leaving localized quasi-stationary peak appeared,  decaying then into a set of wave packets following each other. 
At $t\rightarrow\infty$  a total norm of all outgoing wave packets equals to that of incoming wave packet. At $\varepsilon_F\rightarrow\infty$ the lifetime of the  localized quasi-stationary peak tends to infinity, as it was shown in Ref.\cite{LMT2012}. It is possible to estimate the lifetime of this 
 peak at finite $\varepsilon_F$ as $\tau\sim \varepsilon_F/|V'(r_0)|$, where $r_0$ is defined via the equation $E=V(r_0)$.

\section{Localized states in stationary problem.}
As it was mentioned in a previous section, the functions $f$ and $g$ at $m=0$ are irrelevant to the scattering problem. Therefore, it is interesting to investigate whether these functions correspond to any localized states at some energies. The answer is positive. To illustrate this statement, let us consider the equations for $f$ and $g$ at $m=0$ and fixed energy $E$:
\begin{align}\label{fgE}
E f=\vartheta(\varepsilon_F-p){\hat U}_+f\,,\quad
E g=\vartheta(\varepsilon_F-p){\hat U}_0g\,.
\end{align}
In the particular case of the potential \eqref{pot}, these equations reduce to

\begin{align}\label{fgEpot}
&E f(p)=\frac{u_0 a^2}{2}\vartheta(\varepsilon_F-p)\int_0^\infty dk\,k\exp[-a^2(p^2+k^2)/4]\,I_1(a^2pk/2)f(k)\,,\nonumber\\
&E g(p)=\frac{u_0a^2}{2}\vartheta(\varepsilon_F-p)\int_0^\infty dk\,k\exp[-a^2(p^2+k^2)/4]\,I_0(a^2pk/2)g(k)\,,
\end{align}
where $I_n(x)$ is the modified Bessel function of the first kind. It is seen that $E$ can be written as $E=u_0{\cal E}$, where ${\cal E}$ depends on $\varepsilon_F a$ and is independent of $u_0$.
Passing from the functions $f(p)$ and $g(p)$ to the functions $\sqrt{p}f(p)$ and $\sqrt{p}g(p)$, we obtain the integral equations with the symmetric kernels. Thus, it follows from the theory of such equations that there are infinite set of normalized orthogonal solutions with the energies $|E_1|>|E_2|>|E_3|...$

Let us consider two limiting cases, $\varepsilon_F a\ll 1$ and 
$\varepsilon_F a \gg 1$. In the first case we have $|E_{n+1}/E_n|\ll 1$, and the asymptotic forms of the  solutions for the functions $f$ and $g$ with the maximal energies read
\begin{align}
& f(p)\propto p\vartheta(\varepsilon_F-p)\,, \quad
f(r)=\int_0^\infty dp p J_1(pr)f(p)=\dfrac{1}{r}J_2(\varepsilon_Fr)\,,\quad E_1= u_0(\varepsilon_F a)^4/32\,;\nonumber\\
&g(p)\propto \vartheta(\varepsilon_F-p)\,, \quad
g(r)=\int_0^\infty dp p J_0(pr)g(p)=\dfrac{1}{r}J_1(\varepsilon_Fr)\,,\quad E_1= u_0(\varepsilon_F a)^2/4\,.
\end{align}
In the second case we have $|1-E_{n+1}/E_n|\ll 1$, and the asymptotic forms of the  solutions for the functions $f$ and $g$ with the maximal energies read
\begin{align}
&f(p)\propto \vartheta(\varepsilon_F-p)\,, \quad
f(r)=\frac{1}{\varepsilon_Fr^2}\int_0^{\varepsilon_Fr} dx\, x J_1(x)\,,\quad E_1= u_0\,;\nonumber\\
&g(p)\propto \vartheta(\varepsilon_F-p)\,, \quad
g(r)=\dfrac{1}{r}J_1(\varepsilon_Fr)\,,\quad E_1= u_0\,.
\end{align}
In both cases a typical size of the wave functions is $r\sim 1/\varepsilon_F$.

For $m\neq 0$ the localized solutions with some energies exist only for the function $g$. We have
\begin{align}
 g(p)\propto p^m\vartheta(\varepsilon_F-p)\,, \quad
g(r)=\dfrac{1}{r}J_{m+1}(\varepsilon_Fr)\,,\quad E_1= u_0(\varepsilon_F a/2)^{2m+2}/(m+1)!\,
\end{align}
for $\varepsilon_F a\ll 1$ and
\begin{align}
 g(p)\propto \vartheta(\varepsilon_F-p)\,, \quad
g(r)=\frac{1}{\varepsilon_Fr^2}\int_0^{\varepsilon_Fr} dx\, x J_m(x)\,,\quad E_1= u_0
\end{align}
for $\varepsilon_F a\gg 1$.

\section{Conclusion}
Using the Bethe-Salpeter equation with the kernel calculated in the leading approximation we have derived the wave equation \eqref{DS3} describing the interaction of two electrons in graphene at arbitrary value of the Fermi energy $E_F$ and the equation  \eqref{eqpsi} for the case $E_F\le 0$. 
We have found the explicit forms of the density $\rho(t,\bm r)$ and the current $\bm J_{tot}(t,\bm r)$ which obey the continuity equation \eqref{eqcon}. We have traced how the picture of the wave packet scattering depends on $E_F$. At $m\neq 0$, $E_F<0$, and $E<u_0$, the initial wave comes to the nonzero potential region and the long-leaving localized quasi-stationary peak appears. Then this peak  decays   into a set of wave packets following each other. At $t\rightarrow\infty$  a total norm of all outgoing wave packets equals to that of the incoming wave packet. At $E_F\rightarrow-\infty$ the lifetime of the  localized quasi-stationary peak tends to infinity, which is in agreement with the results of Ref.~\cite{LMT2012} At $E>u_0$ there is only one  outgoing wave with the same shape as the incoming wave  and with the same norm.  At $m= 0$ and any $E_F$, the localized state does not appear in a scattering process. At $E_F=0$ the localized state does not appear for any $m$.

For $E_F<0$ there is an infinite set of the localized solutions with the discrete energies and the finite  norms. A typical size of the localization is $1/|E_F|$. These solutions are irrelevant to the scattering problem. The experimental observation of these states would be a very interesting task.

\bibliographystyle{apsrev}
%\bibliography{Refs}

\begin{thebibliography}{29}
\expandafter\ifx\csname natexlab\endcsname\relax\def\natexlab#1{#1}\fi
\expandafter\ifx\csname bibnamefont\endcsname\relax
  \def\bibnamefont#1{#1}\fi
\expandafter\ifx\csname bibfnamefont\endcsname\relax
  \def\bibfnamefont#1{#1}\fi
\expandafter\ifx\csname citenamefont\endcsname\relax
  \def\citenamefont#1{#1}\fi
\expandafter\ifx\csname url\endcsname\relax
  \def\url#1{\texttt{#1}}\fi
\expandafter\ifx\csname urlprefix\endcsname\relax\def\urlprefix{URL }\fi
\providecommand{\bibinfo}[2]{#2}
\providecommand{\eprint}[2][]{\url{#2}}



\bibitem[{\citenamefont{Novoselov et~al.}(2004)\citenamefont{Novoselov, Geim,
  Morozov, Jiang, Zhang, Dubonos, Grigorieva, and Firsov}}]{Novoselov2004}
\bibinfo{author}{\bibfnamefont{K.~S.} \bibnamefont{Novoselov}},
  \bibinfo{author}{\bibfnamefont{A.~K.} \bibnamefont{Geim}},
  \bibinfo{author}{\bibfnamefont{S.~V.} \bibnamefont{Morozov}},
  \bibinfo{author}{\bibfnamefont{D.}~\bibnamefont{Jiang}},
  \bibinfo{author}{\bibfnamefont{Y.}~\bibnamefont{Zhang}},
  \bibinfo{author}{\bibfnamefont{S.~V.} \bibnamefont{Dubonos}},
  \bibinfo{author}{\bibfnamefont{I.~V.} \bibnamefont{Grigorieva}},
  \bibnamefont{and} \bibinfo{author}{\bibfnamefont{A.~A.}
  \bibnamefont{Firsov}}, \bibinfo{journal}{Science}
  \textbf{\bibinfo{volume}{306}}, \bibinfo{pages}{666} (\bibinfo{year}{2004}).


\bibitem[{\citenamefont{Sabio et~al.}(2010)\citenamefont{Sabio, Sols, and
  Guinea}}]{Sabio2010}
\bibinfo{author}{\bibfnamefont{J.}~\bibnamefont{Sabio}},
  \bibinfo{author}{\bibfnamefont{F.}~\bibnamefont{Sols}}, \bibnamefont{and}
  \bibinfo{author}{\bibfnamefont{F.}~\bibnamefont{Guinea}},
  \bibinfo{journal}{Phys. Rev. B} \textbf{\bibinfo{volume}{81}},
  \bibinfo{pages}{045428} (\bibinfo{year}{2010}).
  
  \bibitem{LMT2012} R. N. Lee, A. I. Milstein, and I. S. Terekhov; Phys. Rev.  B {\bf 86}, 035425 (2012).

\bibitem[{\citenamefont{{Kotov} et~al.}(2010)\citenamefont{{Kotov}, {Uchoa},
  {Pereira}, {Guinea}, and {Castro Neto}}}]{Kotov2011}
\bibinfo{author}{\bibfnamefont{V.~N.} \bibnamefont{{Kotov}}},
  \bibinfo{author}{\bibfnamefont{B.}~\bibnamefont{{Uchoa}}},
  \bibinfo{author}{\bibfnamefont{V.~M.} \bibnamefont{{Pereira}}},
  \bibinfo{author}{\bibfnamefont{F.}~\bibnamefont{{Guinea}}}, \bibnamefont{and}
  \bibinfo{author}{\bibfnamefont{A.~H.} \bibnamefont{{Castro Neto}}},
  \bibinfo{journal}{Rev. Mod. Phys.} \textbf{\bibinfo{volume}{84}},  \bibinfo{pages}{1067} (\bibinfo{year}{2012}).

\bibitem[{\citenamefont{Das~Sarma et~al.}(2011)\citenamefont{Das~Sarma, Adam,
  Hwang, and Rossi}}]{DasSarma2011}
\bibinfo{author}{\bibfnamefont{S.}~\bibnamefont{Das~Sarma}},
  \bibinfo{author}{\bibfnamefont{S.}~\bibnamefont{Adam}},
  \bibinfo{author}{\bibfnamefont{E.~H.} \bibnamefont{Hwang}}, \bibnamefont{and}
  \bibinfo{author}{\bibfnamefont{E.}~\bibnamefont{Rossi}},
  \bibinfo{journal}{Rev. Mod. Phys.} \textbf{\bibinfo{volume}{83}},
  \bibinfo{pages}{407}  (\bibinfo{year}{2011}).
  
  \bibitem[{\citenamefont{Wallace}(1947)}]{Wallace1947}
\bibinfo{author}{\bibfnamefont{P.~R.} \bibnamefont{Wallace}},
  \bibinfo{journal}{Phys. Rev.} \textbf{\bibinfo{volume}{71}},
  \bibinfo{pages}{622} (\bibinfo{year}{1947}).

\bibitem[{\citenamefont{McClure}(1956)}]{McClure1956}
\bibinfo{author}{\bibfnamefont{J.~W.} \bibnamefont{McClure}},
  \bibinfo{journal}{Phys. Rev.} \textbf{\bibinfo{volume}{104}},
  \bibinfo{pages}{666} (\bibinfo{year}{1956}).

\bibitem[{\citenamefont{Gonz\'alez et~al.}(1993)\citenamefont{Gonz\'alez,
  Guinea, and Vozmediano}}]{Gonzalez1993}
\bibinfo{author}{\bibfnamefont{J.}~\bibnamefont{Gonz\'alez}},
  \bibinfo{author}{\bibfnamefont{F.}~\bibnamefont{Guinea}}, \bibnamefont{and}
  \bibinfo{author}{\bibfnamefont{M.}~\bibnamefont{Vozmediano}},
  \bibinfo{journal}{Nuclear Physics B} \textbf{\bibinfo{volume}{406}},
  \bibinfo{pages}{771 } (\bibinfo{year}{1993}), ISSN \bibinfo{issn}{0550-3213}.

\bibitem[{\citenamefont{Gonz\'alez et~al.}(1994)\citenamefont{Gonz\'alez,
  Guinea, and Vozmediano}}]{Gonzalez1994}
\bibinfo{author}{\bibfnamefont{J.}~\bibnamefont{Gonz\'alez}},
  \bibinfo{author}{\bibfnamefont{F.}~\bibnamefont{Guinea}}, \bibnamefont{and}
  \bibinfo{author}{\bibfnamefont{M.}~\bibnamefont{Vozmediano}},
  \bibinfo{journal}{Nuclear Physics B} \textbf{\bibinfo{volume}{424}},
  \bibinfo{pages}{595 } (\bibinfo{year}{1994}), ISSN \bibinfo{issn}{0550-3213}.

\bibitem[{\citenamefont{Berestetski et~al.}(1982)\citenamefont{Berestetski,
  Lifshits, and Pitayevsky}}]{Berestetski1982}
\bibinfo{author}{\bibfnamefont{V.}~\bibnamefont{Berestetski}},
  \bibinfo{author}{\bibfnamefont{E.}~\bibnamefont{Lifshits}}, \bibnamefont{and}
  \bibinfo{author}{\bibfnamefont{L.}~\bibnamefont{Pitayevsky}},
  \emph{\bibinfo{title}{Quantum electrodynamics}}
  (\bibinfo{publisher}{Pergamon}, \bibinfo{year}{1982}).
  

\bibitem[{\citenamefont{Itzykson et~al.}(1982)\citenamefont{Itzykson, Zuber}}]{IZ}
\bibinfo{author}{\bibfnamefont{C.}~\bibnamefont{Itzykson}},
\bibnamefont{and}
  \bibinfo{author}{\bibfnamefont{J.-B.}~\bibnamefont{Zuber}},
  \emph{\bibinfo{title}{Quantum field theory}}
  (\bibinfo{publisher}{McGraw-Hill}, \bibinfo{year}{1980}).

%\bibitem{}C. Itzykson and J-B Zuber, \it{Quantum field theory}(...).

\bibitem[{\citenamefont{Gamayun et~al.}(2009)\citenamefont{Gamayun, Gorbar, and
  Gusynin}}]{Gamayun2009}
\bibinfo{author}{\bibfnamefont{O.~V.} \bibnamefont{Gamayun}},
  \bibinfo{author}{\bibfnamefont{E.~V.} \bibnamefont{Gorbar}},
  \bibnamefont{and} \bibinfo{author}{\bibfnamefont{V.~P.}
  \bibnamefont{Gusynin}}, \bibinfo{journal}{Phys. Rev. B}
  \textbf{\bibinfo{volume}{80}}, \bibinfo{pages}{165429}
  (\bibinfo{year}{2009}).



  
  \bibitem[{\citenamefont{Gamayun et~al.}(2010)\citenamefont{Gamayun, Gorbar, and
  Gusynin}}]{Gamayun2010}
\bibinfo{author}{\bibfnamefont{O.~V.} \bibnamefont{Gamayun}},
  \bibinfo{author}{\bibfnamefont{E.~V.} \bibnamefont{Gorbar}},
  \bibnamefont{and} \bibinfo{author}{\bibfnamefont{V.~P.}
  \bibnamefont{Gusynin}}, \bibinfo{journal}{Phys. Rev. B}
  \textbf{\bibinfo{volume}{81}}, \bibinfo{pages}{075429}
  (\bibinfo{year}{2010}).
  
  
  \bibitem[{\citenamefont{Downing et~al.}(2010)\citenamefont{Downing  and
  Portnoi}}] {port2017}
\bibinfo{author}{\bibfnamefont{C.~A.} \bibnamefont{Downing}} ,
  \bibinfo{author}{\bibfnamefont{M.~E.} \bibnamefont{Portnoi}}, 
  \bibinfo{journal}{Nat. Comm.}
  \textbf{\bibinfo{volume}{8}}, \bibinfo{pages}{897}
  (\bibinfo{year}{2017}).
  



\end{thebibliography}

\end{document}